\documentstyle[aps,epsfig]{revtex}
\begin{document}
\begin{titlepage}
\pagestyle{empty}
\baselineskip=18pt
\rightline{NBI--97--07}
\rightline{March, 1997}
\baselineskip=15pt
\vskip .2in
\begin{center}
{\large{\bf Fluctuations and HBT Scales 
in Relativistic Nuclear Collisions}}
\end{center}
\vskip .2truecm
\begin{center}
Henning Heiselberg 

{\it NORDITA, Blegdamsvej 17, DK-2100 Copenhagen \O., Denmark}

and Axel P.~Vischer

{\it Niels Bohr Institute, DK-2100, Copenhagen \O, Denmark.}

\end{center}

\vskip 0.1in
\centerline{ {\bf Abstract} }
\baselineskip=15pt
\vskip 0.5truecm

Bose-Einstein correlations in relativistic heavy ion collisions are
examined in a general model containing the essential features of
hydrodynamical, cascade as well as other models commonly employed for
describing the particle freeze-out. In particular the effects of
longitudinal and transverse expansion, emission from surfaces moving
in time, the thickness of the emitting layer varying from surface to
volume emission and other effects are studied. Model dependences of
freeze-out sizes and times are discussed and compared to recent
$Pb+Pb$ data at 160A$\cdot$GeV.
 
\end{titlepage} 

\baselineskip=15pt
\textheight8.9in\topmargin-0.0in\oddsidemargin-.0in

\section{Introduction}

Bose-Einstein interference of identical particles or the Hanbury-Brown
\& Twiss effect (HBT) \cite{HBT} shows up in correlation functions of
pions and kaons emitted from the collision zone in relativistic heavy
ion collisions. It is an important tool for determining the source at
freeze-out and recent data \cite{NA44,NA44QM,NA49,WA80,AGS} can
restrict the rather different models, that have been developed to
describe particle emission in high energy nuclear collisions. In
hydrodynamical calculations particles freeze-out at a hypersurface
that generally does not move very much transversally until the very
end of the freeze-out \cite{hydro}.  In cascade codes the last
interaction points are also found to be distributed in transverse
direction around a mean value that does not change much with time
\cite{Humanic,RQMD,QGSM,Scott}, but the width of the emission zone increases
from narrow surface emission to a widespread volume emission.  Other
models like cylindrical symmetric models with Bjorken longitudinal
scaling and volume emission \cite{Csorgo,Heinz} or surface emitting
sources \cite{Pratt,Opaque} have also been studied and sizes,
freeze-out times, etc. have been estimated by comparing to
experimental data.

In this letter we want address the general dynamical features of
particle emission in relativistic heavy ion collisions. We start from 
a general source, incorporating volume as well as surface emission, 
longitudinal and transverse flow and expansion as well as moving
freeze-out surfaces. The various contributions to the HBT radius
parameters are calculated and we show, that
they can be interpreted as fluctuations in radial, temporal, angular and
emission layer thickness variables. Finally, we discuss a more
quantitative analysis and compare to recent NA44 HBT data on central
$Pb+Pb$ collisions at $160$A$\cdot$GeV.

For the correlation function analysis of Bose-Einstein interference
from a source of size $R$ we consider two particles emitted a distance
$\sim R$ apart with relative momentum ${\bf q}=({\bf k}_1-{\bf k}_2)$
and average momentum, ${\bf K}=({\bf k}_1+{\bf k}_2)/2$. Typical heavy
ion sources in nuclear collisions are of size $R\sim5$ fm, so that
interference occurs predominantly when 
$q\raisebox{-.5ex}{$\stackrel{<}{\sim}$}\hbar/R\sim 40$
MeV/c. Since typical particle momenta are $k_i\simeq K\sim 300$ MeV,
the interfering particles travel almost parallel (see Fig. (1)), i.e.,
$k_1\simeq k_2\simeq K\gg q$.  The correlation function due to
Bose-Einstein interference of identical particles from an incoherent
source is (see, e.g., \cite{Heinz})
\begin{equation}
   C_2({\bf q},{\bf K})=1\;\pm\; |\frac{\int d^4x\;S(x,{\bf K})\;e^{iqx}}
   {\int d^4x\;S(x,{\bf K})}|^2 \,, \label{C}
\end{equation}
where $S(x,{\bf K})$ is a function describing the phase space density of the
emitting source. The $+/-$ refers to boson/fermions respectively.

Experimentally the correlation functions for identical mesons
($\pi^\pm\pi^\pm$, $K^\pm K^\pm$, etc.) are often
parametrized by the gaussian form
\begin{equation}
  C_2(q_s,q_o,q_l)=1+\lambda\exp(-q_s^2R_s^2-q_o^2R_o^2-q_l^2R_l^2
  -2q_oq_lR_{ol}^2  )\;.   \label{Cexp}
\end{equation}
Here, ${\bf q}={\bf k}_1-{\bf k}_2=(q_s,q_o,q_l)$ is the relative momentum
between the two particles and $R_i,i=s,o,l$ the corresponding sideward,
outward and longitudinal HBT radius parameters respectively.
We will employ the standard geometry where the {\it longitudinal} direction
is along the beam axis and the outward direction is along ${\bf K}$ and
the sideward axis is perpendicular to these.
Usually, each pair of mesons are lorenzt boosting longitudinal to the system 
where their longitudinal momentum or rapidity vanish, $Y=0$; here 
their average momentum
${\bf K}$ is perpendicular to the beam axis and is chosen as the {\it outward}
direction. In this system the pair velocity
\mbox{\boldmath $\beta_K$}=${\bf K}/E_K$ points in
the outward direction with $\beta_o=p_\perp/m_\perp$ where 
$m_\perp=\sqrt{m^2+p_\perp^2}$ is the transverse mass.
As pointed out in \cite{Heinz} the out-longitudinal coupling $R_{ol}$
only vanish to leading order when $Y=0$.
The reduction factor $\lambda$ in Eq. (\ref{Cexp}) may be
due to long lived resonances \cite{HH,Csorgo}, coherence effects,
incorrect Gamov corrections \cite{BB} or other effects. It is found to
be $\lambda\sim 0.5$ for pions and $\lambda\sim 0.9$ for kaons.

\begin{figure}
\centerline{
\psfig{figure=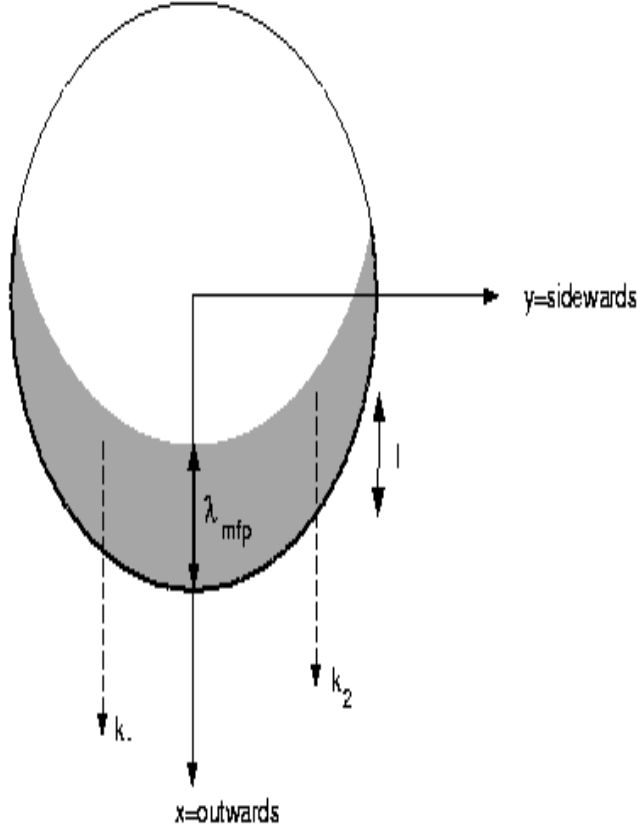,width=13cm,height=9.5cm,angle=-90}}
\caption{Cross section of the interaction region perpendicular to the
longitudinal or z--direction. Particles have to penetrate a distance
$l$ out to the surface of the interaction region in order to escape and reach
the detector. The bulk part of the emitted particles comes from a surface
region of width $\lambda_{mfp}$, the mean free path of the particle.
}
\end{figure}

It is very convenient to introduce the the source average and fluctuation
or variance of a quantity ${\cal O}$ defined by
\begin{equation}
 \langle{\cal O}\rangle\equiv
   \frac{\int d^4x\;S(x,{\bf K}){\cal O}}{\int d^4x\;S(x,{\bf K})}\;, \quad
 \sigma({\cal O}) \equiv \langle{\cal O}^2\rangle -\langle{\cal O}\rangle^2
    \;. \label{O}
\end{equation} 
With $qx\simeq{\bf q\cdot x}-{\bf q\cdot}$\mbox{\boldmath $\beta$}$_K\,t$
one can, by expanding to second order in $q_i\;R_i$ and compare to
Eq. (\ref{Cexp}), find the HBT radius parameters $R_i$, {\it i=s,o,l}.
\footnote[1]{The expansion at small $q_i$ and extraction of the HBT
radius parameters $R_i$, Eq. (\ref{Ri}), may not be directly
comparable to the gaussian radii extracted experimentally because the
experimental data has best statistics around $q_i\sim \hbar/R_i\sim
40$MeV/c. For the specific models discussed here we have checked that
they are comparable. We attribute this agreement to the many different
fluctuations contributing to the HBT radius parameters which ``smear''
out the source; for many random fluctuations we expect that the
central-limit theorem guaranties a gaussian distribution.}  They are
(see, e.g. \cite{Heinz})
\begin{equation}
   R_i^2=\sigma(x_i-\beta_i\;t) \,. \label{Ri}
\end{equation}
The HBT radius parameters are a measure for the fluctuations of
$(x_i-\beta_it)$ over the source emission function $S$.

  Particle production in ultrarelativistic heavy ion collisions has
been found to have dynamical features of strong longitudinal expansion
and some transverse expansion \cite{NA44slopes}.  Furthermore, the
geometry plays a role and one would expect that particles escape from
the outer layers of the fireball or if the source is sufficiently
small the whole volume will freeze-out.  We model such a general
cylindrically symmetric source by assuming local thermal equilibrium
with longitudinal Bjorken flow ($u_z={\rm z}/t$) as well as transverse
flow ${\bf v}$ through a Boltzmann factor.  The space-time geometry is
modeled through the transverse source distribution $S_\perp(r_\perp)$
and the temporal part $S_\tau(\tau)$
\begin{equation}
   S(x,{\bf K}) \sim e^{-K\cdot u/T}\,
          S_\perp(r_\perp)\, S_\tau(\tau) \,. \label{S}
\end{equation}
Here, $\tau=\sqrt{t^2-{\rm z}^2}$ is the invariant
time and $\eta= 0.5\ln(t+{\rm z})/(t-{\rm z})$ the space-time rapidity.
Including transverse flow ${\bf v}=(v_x,v_y)$ 
the flow four-vector is 
$u=\gamma(v)(\cosh(\eta),\sinh(\eta),{\bf v})$ \cite{Csorgo,Heinz},
which gives 
\begin{eqnarray}
   K\cdot u = m_\perp\gamma(v)(\cosh(\eta-Y)-{\bf \beta}_o\cdot{\bf v})\,.
     \label{Ku}
\end{eqnarray}
 Here $m_\perp$ is the transverse mass and $Y$ the rapidity of the
particles. In the following all variables are boosted into the
frame in which $Y=0$.  Notice that any normalization cancels out
in correlation function (\ref{C}).  One can in addition apply a factor
$S_\eta(\eta)\sim\exp(-(\eta-\eta_{cms})^2/2\delta\eta^2)$ to correct
for the lack of Bjorken scaling near target and projectile rapidities
\cite{Csorgo}.  However, the thermal factor
$\exp(-m_\perp\cosh(\eta-Y)/T)$ centers the space-time rapidity $\eta$
around the pair rapidity $Y$ on a scale $\sim\sqrt{T/m_\perp}$ much
narrower than $\delta\eta$. The additional factor has thus only a
minor effect, reducing the longitudinal source sizes slightly.

It is important to distinguish between volume and surface freeze--out,
since they give very different HBT radius parameters~\cite{Opaque}.
Hydrodynamical models~\cite{hydro} assume that particles are emitted
from the surface. This is actually also found in some cascade models
at early times of the collision~\cite{Humanic,RQMD,QGSM,Scott}, but
eventually the whole source freezes out and disintegrates. The late
stage of cascade models resembles more a volume freeze--out. We want
to mimic these different models and stages by introducing a source
which emits particles according to a simple Glauber theory. The
surface freeze--out component is described by a Glauber absorption
factor, which suppresses particles escaping from the interior of the
source. The source becomes opaque
\begin{equation}
   S_\perp(r_\perp) \sim \, e^{-\int_x^\infty dx'\sigma\rho(x')}
      \,, \label{SGlauber}
\end{equation}
where the integral runs over the particle trajectory from last interaction
point $x$. Modifications of
single particle spectra in hydrodynamic calculations due to such an
emission layer (\ref{SGlauber}) has been consider in \cite{Grassi}.
We introduce the mean free path 
$\lambda_{mfp}=(\sigma\langle\rho\rangle)^{-1}$,
where $\langle\rho\rangle$ is the average density in the 
emission layer and $\sigma$ the interaction cross section.
Glauber absorption only allows emission from a 
layer of thickness $\sim\lambda_{mfp}$ just inside the surface radius $R$.
Thus we can rewrite
\begin{equation}
   S_\perp(r_\perp) \sim \, e^{-l/\lambda_{mfp}}\Theta(R-r_\perp)\,,
        \label{Glauber}
\end{equation}
where $l=\sqrt{R^2-{\rm y}^2}-{\rm x}$ is the distance the particle
has to pass through the source in order to escape from the surface
when ${\rm (x,y)}=R\, (\cos\theta,\sin\theta)$ is its position in outward
and sideward direction respectively (see Fig. (1)). In hydrodynamical
calculations $\lambda_{mfp}=0$ but in cascade codes it can be several
{\it fm}'s .  The surface may also move in the transverse direction
with time, $R(\tau)$.

The temporal emission of the source is determined by
$S_\tau(\tau)$. It is commonly approximated by a gaussian,
$S_\tau(\tau)\sim \exp(-(\tau-\tau_0)^2/2\delta\tau^2)$, around the
source mean life-time, $\tau_0$ with width, $\delta\tau$, which is the
duration of emission.  These gaussian parameters approximate the
average emission time, $\langle\tau\rangle$ and the variance or
fluctuation, $\sigma(\tau)$, respectively for a general source.

An opaque source emits from a thin surface layer instead from the whole
volume. We can thus calculate the HBT radius parameters very generally by
expanding in $\lambda_{mfp}/R\ll 1$. For strict surface emission,
$\lambda_{mfp}\simeq 0$ and the source (\ref{Glauber}) reduces to
\begin{equation}
  S_\perp \sim \delta(R(\tau)-r_\perp)\Theta(\cos\theta)\,\cos\theta \,. 
   \label{Ssurf}
\end{equation}
 The geometric factor $\cos\theta$ suppress the peripheral zones and
the $\Theta(\cos\theta)$ factor insures that
particles are only emitted {\it away} from the surface, i.e., only
particles from the surface layer of the half hemisphere directed
towards the detector will reach it whereas particles from the other
hemisphere will interact on their passage through the
source.\footnote[2]{Detectors on the other side of the beam line measure
particle from the other hemisphere. Thus relativistic heavy ion
collisions have an advantage to stellar interferometry which cannot
measure the back side or, as referred to in the case of the moon, the
``dark side''.}  Most hydrodynamical freeze-out mechanisms do not
include the directional condition that particles can only be emitted
away from the surface though strong flow at the surface has a similar
effect. Whereas the source in (\ref{Ssurf}) corresponds to black body
emission, which is constant in time per surface element, the temporal
variation is described by $S_\tau(\tau)$.

The HBT radius parameters can now be calculated from Eq. (\ref{O}) using Eqs.  (\ref{S})
and (\ref{Glauber}) for any mean free path and 
transverse flow and results are shown in Fig. 2.
It is, however, very instructive to consider the case of emission from
a surface layer, $\lambda_{mfp}\ll R(\tau)$, where a number of
simplifications appear and analytical results can be obtained. 
Since the emission points are narrowly confined within a layer
of thickness $\lambda_{mfp}$ within the surface, 
an expansion in the mean free path can be performed
and the integration in transverse radial direction simplifies.
The space-time rapidity, angular and temporal integrations separates and
due to the normalization a number of factors cancel when evaluating the
HBT radius parameters. For example, a function of proper time only needs to be
averaged with respect to the temporal parts of the source 
\begin{eqnarray}
   \langle {\cal O}(\tau)\rangle=\frac{\int^{\tau_f}_0d\tau\,\tau R(\tau)\,
           S_\tau(\tau) {\cal O}(\tau)}
           {\int^{\tau_f}_0d\tau\,\tau R(\tau)S_\tau(\tau)} \,. \label{tau}
\end{eqnarray}
The angular averages also simplify for the cylindrical geometry.
From the definitions in Eq. (\ref{O}) we obtain
\begin{eqnarray}
 \langle {\cal O}(\theta)\rangle 
 &=& \frac{\int_{-\pi/2}^{\pi/2} {\cal O}(\theta) 
     \exp(\gamma(v_s)v_s p_\perp\cos\theta/T)\cos\theta\, d\theta }
   {\int_{-\pi/2}^{\pi/2} \exp(\gamma(v_s)v_s p_\perp\cos\theta/T)\cos\theta\, 
     d\theta }      \,. \label{Theta}
\end{eqnarray}  
The factor $\exp(\gamma(v_s)v_s p_\perp\cos\theta/T)$ includes the
effect of transverse flow $v_s$ at the surface (see Eq. (\ref{Ku}) and
has the effect of narrowing the angular emission in the direction of
${\bf K}$, i.e. in the outward direction ($\theta=0$). A spherical
source would have an additional factor $R(\tau)$ in (\ref{tau}) and
$\sin\theta$ in the integrals of Eq. (\ref{Theta}).

The HBT radius parameters can to leading order in fluctuations
and to second order in $\lambda_{mfp}\ll R(\tau)$ (in the frame where
$Y=0$) now be evaluated $^{\ddag}$
\begin{eqnarray}
   R_s^2 &\equiv& \sigma({\rm y}) = \langle {\rm y}^2\rangle 
         =  \langle R(\tau)^2\rangle \sigma(\sin\theta)
            \,-\,\frac{1}{6} \lambda_{mfp}^2  \,,\label{Rsf}\\
   R_o^2 &\equiv& \sigma({\rm x}-\beta_o t)  \nonumber\\
         &=& \langle R(\tau)\rangle^2 \sigma(\cos\theta)
             \,+\, \sigma(R(\tau))\langle\cos^2\theta\rangle
             \,+\, \beta_o^2 \sigma(\tau) \nonumber\\
         &&    -\, 2\beta_o\langle\cos\theta\rangle 
             \langle(R(\tau)-\langle R(\tau)\rangle)
            (\tau-\langle\tau\rangle)\rangle 
             \,+\, (\frac{7}{6}-\frac{\pi^2}{32}) \lambda_{mfp}^2  
              \,, \label{Rof}\\
   R_l^2 &\equiv& \sigma(z-\beta_l t) 
          = \sigma(\tau\sinh\eta)
          \simeq  \frac{\langle\tau^2\rangle}{\gamma(v_s)}\frac{T}{m_\perp} 
          \,. \label{Rlf}
\end{eqnarray}

The terms in $R_s^2$ are the average of the square of the transverse
source size times the angular fluctuations which, due to the inversion
symmetry along the y-axis, is
$\sigma(\sin\theta)=\langle\sin^2\theta\rangle$.\footnote[3]{For a
cylindrically symmetric source without transverse flow \cite{Opaque}
$\sigma(\sin\theta)=\langle\sin^2\theta\rangle=1/3$,
$\langle\cos^2\theta\rangle=2/3$, and
$\sigma(\cos\theta)=2/3-(\pi/4)^2\simeq 0.05$. In the coefficients to
$\lambda_{mfp}^2$ we have, for simplicity, neglected transverse
flow.}  As the emission layer is inside the source particle have
$r_\perp<R$ and $R_s$ the sideward HBT radius parameter is reduced by a term of
order$^{\ddag}$ $\lambda_{mfp}^2$.

The terms in $R_o^2$ are respectively:
{\it i)} angular fluctuations$^{\ddag}$:
$\sigma(\cos\theta)=\langle\cos^2\theta\rangle-\langle\cos\theta\rangle^2$, 
{\it ii)} fluctuations in transverse radial direction: 
$\sigma(R(\tau))= \langle R^2(\tau)\rangle -\langle R(\tau)\rangle^2$,
{\it iii)} temporal fluctuations: 
$\sigma(\tau)=\langle\tau^2\rangle -\langle\tau\rangle^2$,
{\it iv)} a cross term between radial and temporal variations 
which is {\it positive} for an {\it inward} moving surface,
{\it v)}  and finally a term of order $\lambda_{mfp}^2$ due to
the thickness of the surface layer {\it adding} to the outward fluctuations.
In addition the average life-times and fluctuations in the life-times of 
short lived resonances should be added to the temporal fluctuations
\cite{HH}; the long lived resonances can account for most of the
reduction factor $\lambda$ in Eq. (\ref{Cexp}).

It was assumed that $Y=0$, i.e. the longitudinal pair velocity
vanishes $\beta_l=0$.
The longitudinal HBT radius parameter is then the fluctuation 
$\sigma(z)=\sigma(\tau\sinh\eta)$. When $T\ll m_\perp$ the
space-time rapidity $\eta$ is small and
one finds $\sigma(\sinh\eta)\simeq\sigma(\eta)\simeq T/m_\perp$.

A transparent source corresponds to the opposite limit of
(\ref{Glauber}), i.e. $\lambda_{mfp}\rightarrow\infty$.  The
transverse radius parameters are in this case the same $R_s=R_o=R/2$
for a source without transverse flow or fluctuations in (duration of)
emission time. The longitudinal HBT radius parameter is unchanged for
$m_\perp\gg T$, since it is only sensitive to the longitudinal
expansion and emission time but not to transverse coordinates.  The
general dependence on transparency/opacity is shown in Fig. (2),
where the transverse radius parameters are plotted as functions of the
transverse flow parameter $\gamma v_s p_\perp/T$ for different
$\lambda_{mfp}/R$.  A flow profile of the form ${\bf v}=v_s\;{\bf
r}/R_0$ was assumed. For simplicity we have neglected the radial
dependence hidden in $\gamma(v_s)$, which is allowed for small
mean free paths or small transverse flow velocities. For transparent
sources it reduces the transverse HBT radius parameters by the same amount to
order $v_s^2$ \cite{Csorgo}.  Also notice that only spatial
fluctuations are included in $R_0^2$ whereas temporal fluctuations
such as $\beta_o^2\, \sigma(\tau)$ still have to be added.

\begin{figure}
\centerline{
\psfig{figure=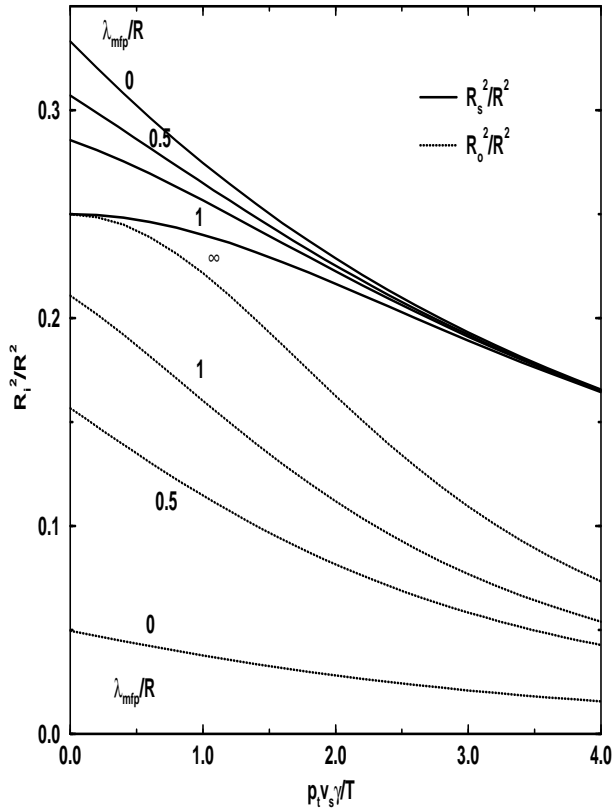,width=13cm,height=9.5cm,angle=-90}}
\caption{
Sideward and outward radii as function of the transverse flow
parameter $\gamma v_s p_\perp/T$. The curves correspond to different
mean free paths as labeled (see text).
}
\end{figure}

For a transparent source with $\lambda_{mfp}\gg R$ and without transverse
flow the outward HBT radius parameter is larger than the sideward
\cite{Csorgo,Heinz}
\begin{eqnarray}
   R_o^2=R_s^2+\beta_o^2\sigma(\tau) \, . \label{R1}
\end{eqnarray} 
The excess is due to the duration of emission, $\sigma(\tau)^{1/2}$,
of the source in which particles with outward velocity
$\beta_o=p_\perp/m_\perp$ travel on average a distance
$\beta_o\, \sigma(\tau)^{1/2}$ towards the detector. The sideward
distance is perpendicular to this velocity and $R_s$ is therefore
affected by the duration of emission and thus reflects the ``true''
transverse size of the source. The longitudinal HBT radius parameter $R_l$ is
affected by the duration of emission through the factor
$\langle\tau^2\rangle=\langle\tau\rangle^2+\sigma(\tau)$.

In relativistic heavy ion collisions
the outward and sideward HBT radius parameters are measured to be similar
\cite{NA44,NA44QM,NA49,WA80,AGS} and in a few cases the outward is
even measured to be smaller than the sideward HBT radius parameter
\cite{NA44,NA44QM} contradicting Eq. (\ref{R1}), however, within
experimental uncertainty.  According to Eq. (\ref{R1}) this implies
that particles freeze-out suddenly, $\delta\tau\ll R_i$, as in a
``flash'' \cite{CC}, in particular when resonance life-times are included
\cite{HH,Csorgo}.  However, both the opacity effect leading to surface
emission and transverse flow reduces $R_o$ more than $R_s$ and so it
is possible that $R_o<R_s$. The opacity effect is evident from
Eqs. (\ref{Rsf}) and (\ref{Rof}) and is due to the surface emission
which narrows the emission to $\langle{\rm
x}^2\rangle\simeq\langle{\rm x}\rangle^2$ (see Fig. (1)) which reduces
the fluctuations $\sigma({\rm x})$ significantly. Transverse flow
narrows the emission along ${\bf K}$, i.e. in the outward direction,
as can be seen from Eq. (\ref{Theta}) and Fig. 2 which therefore
reduces $R_s$ more than $R_o$. For a source with surface emission
and/or transverse flow it is therefore possible to obtain $R_o<R_s$
but a number of other effects will add fluctuations such as 
$\sigma(\tau), \sigma(R(\tau), \lambda_{mfp}^2$ and more
will add on $R_o^2$  (see Eq. (\ref{Rom}).

To investigate the possibility $R_o<R_s$ it is instructive to consider
a source similar to what is found in hydrodynamic models. The
freeze-out surface remains at an almost constant transverse distance
$R$ initially but eventually moves rapidly inwards at final time
$\tau_f$. More specifically we choose $R(\tau)={\cal
R}\sqrt{1-\tau^2/\tau_f^2}$ and constant particle emission per surface
element ($S_\tau(\tau)=constant$). Inserting the source in
Eqs. (\ref{Rsf}-\ref{Rlf}) we find without transverse flow ($v=0$)
\begin{eqnarray}
   R_s^2 &=& \frac{1}{5}{\cal R}^2 \, -\, \frac{1}{6}\lambda_{mfp}^2  
             \,,\label{Rsm}\\
   R_o^2 &=& c_1{\cal R}^2 + c_1\beta_o^2 \tau_f^2 +c_2\beta_o\tau_f {\cal R}
          \,+\,(\frac{7}{6}-\frac{\pi^2}{32}) \lambda_{mfp}^2   
           \,. \label{Rom}\\
   R_l^2 &=& \frac{2}{5}\frac{T}{m_\perp} \tau_f^2 \,. \label{Rlm}
\end{eqnarray}
where $c_1=2/5-(3\pi/16)^2\simeq0.053$ and
$c_2=2(3\pi/16)^2-\pi/5\simeq0.066$. The terms in $R_o^2$ correspond
to the terms in Eq. (\ref{Rof}) from angular and radial fluctuations,
the temporal fluctuations, the cross term, and finally the
fluctuations from the width of the surface layer respectively.

In the recent NA44 data on central $Pb+Pb$ collisions at 160
A$\cdot$GeV pion HBT radius parameters of $R_s\simeq R_o\sim 4.5-5.0$
fm and $R_l\simeq5-6$ fm are found \cite{NA44}. The average transverse
momentum was $p_\perp\simeq 165$ MeV such that $\beta_o\simeq 0.76$.
>From the transverse momentum slopes of pions, kaon, protons and
deuterium in \cite{NA44slopes} one finds a temperature $T\sim 120$ MeV
and transverse flow $\langle v_s^2 \rangle^{(1/2)} \sim 0.6c$. The
transverse flow effects on the HBT radius parameters (see equation
(\ref{Theta}) and Fig. 2 with $\gamma v_s p_\perp/T\sim 1$) are small
for these numbers which allows us to use the HBT radius parameters
determined in equations (\ref{Rsm}-\ref{Rlm}).  As seen from equation
(\ref{Rsm}) and (\ref{Rom}) and Fig. 2 the sideward radius is
relatively less affected by the $\lambda_{mfp}^2$ correction than the
outward.  We extract a freeze-out time $\tau_f\sim 12$ fm/c and an
initial transverse source size ${\cal R}\sim 11$ fm from the
experimental $R_l$ and $R_s$ respectively.  From the experimental
value for $R_o$ we finally extract $\lambda_{mfp}\sim2-3$ fm.
The initial transverse source size is larger than the
geometrical size of $Pb$ $\sim7$ fm by an amount in excess of the
uncertainty in the impact parameter, expected from the centrality
cuts. Some expansion of the source seems to have taken place before
final freeze-out. The average emission time is
$\sqrt{\langle\tau^2\rangle}=\sqrt{2/5}\tau_f\simeq 7$ fm/c which is
entirely consistent with the freeze-out time required to explain the
enhancement in the $\pi^-/\pi^+$ ratio at low $p_\perp$ due to Coulomb
repulsion in the same $Pb+Pb$ collisions \cite{Barz}.

In the above example the pions are emitted during the whole period
from collision to freeze-out and do not appear as in a ``flash''. The
opacity effect is more important in reducing $R_o$ than transverse
flow at this $p_\perp$. Fluctuations from radial, angular, temporal,
cross term and thickness of emission layer contribute by similar
amounts to $R_o$. At large transverse momenta the sideward and outward
HBT radius parameters are reduced by transverse flow and the
longitudinal radius parameter scales like $1/m_\perp$.  These results
are in qualitative agreement with the experiment \cite{NA44}.

Eqs. (\ref{Rsm}-\ref{Rom}) leads to $R_o<R_s$ for small $\beta_o$ and
$\lambda_{mfp}$.  It is, however, crucial to realize that this simple
model breaks down when the inward moving surface speed exceeds the
particle velocity outwards \cite{Scott}. When that occurs the
particles are overtaken by the surface and experience a volume
freeze-out rather than the Glauber picture employed above in which the
particles scatter their way out through the surface.  Near freeze-out
the surface moves rapidly inwards and Eqs. (\ref{Rsm}-\ref{Rlm}) are
therefore not valid - especially at low $\beta_o=p_\perp/m_\perp$.

As mentioned cascade codes results in sources that can approximately be
described by two components, initially surface emission but eventually
volume freeze-out. Generally, for a two-component source
\begin{eqnarray}
  S(x) = p\, S_1(x) + (1-p)\, S_2(x) \,, \label{2comp}
\end{eqnarray}
with normalization $\int d^4x S_i(x)=1$ such that $p$ is the fraction of
particles from source 1,
the fluctuations in a quantity ${\cal O}$ is by inserting (\ref{2comp})
in (\ref{O}) 
\begin{eqnarray}
  \sigma({\cal O}) = p\,\sigma_1({\cal O}) + (1-p)\,\sigma_2({\cal O})
   +p\,(1-p)\,(\langle{\cal O}\rangle_1-\langle{\cal O}\rangle_2)^2 \,.
 \label{p}
\end{eqnarray}
Here, $\langle{\cal O}\rangle_i\equiv \int d^4x S_i(x){\cal O}$ and
$\sigma_i({\cal O})=\langle{\cal O}^2\rangle_i-\langle{\cal
O}\rangle_i^2$. The fluctuations are the weighted sum of the
fluctuations in the individual sources and an additional cross term.
Since $\langle {\rm y}\rangle=0$ and $\langle z-\beta_l t\rangle$ also
vanishes for $Y=0$ this additional cross term does not contribute to
the sideward and longitudinal HBT radius parameters. It is, however,
nonnegligible for the outward HBT radius parameter.

 The sideward and longitudinal HBT radius parameters are less sensitive to the
details of the freeze-out than the outward HBT radius parameter, which
allows us to estimate the transverse source size and freeze-out.  For
example, for a source emitting from a static surface we find $R_s={\cal
R}/\sqrt{3}$ \cite{Opaque}, for volume emission of constant density we
obtain
$R_s={\cal R}/2$ (see above) and for an inward moving source $R_s$ is
given by Eq. (\ref{Rsm}).  These expressions give a transverse source
size in the range ${\cal R}\sim 9-11$ fm from the NA44 data on
$R_s$. The longitudinal HBT radius parameter depends on
$\sqrt{\langle\tau^2\rangle}$ which is simply $\tau_f$ for a source
emitting particles at freeze-out but smaller by a factor $\sqrt{2/5}$
for constant emission per surface element, Eq. (\ref{Rlm}).  The
freeze-out time thus lies in the range $\tau_f\simeq 8-11$ fm/c using
the NA44 data on $R_l$. In contrast the outward HBT radius parameter is very
model dependent as it contains contributions from a number of unknown
quantities such as radial, angular, temporal and width of the emission
layer, cross terms between radial and temporal correlations as seen
from Eq. (\ref{Rom}) as well as transverse flow. In addition to the
strong dependence on whether the source is opaque or transparent, a
mixture of these sources produces a cross term in $R_o^2$,
Eq. (\ref{p}), besides the individual fluctuations.

In summary we have expressed the HBT radius parameters as averages and
fluctuations in spatial and temporal quantities for a general class of
sources with longitudinal and transverse expansion as well as surface
and volume emission changing with time. The outward HBT radius
parameter $R_o$ consists of fluctuations in radial, angular and
temporal quantities, the width of emission layer as well as life-times
of short lived resonances. Glauber absorption leads to emission from a
surface layer away from the source and has the effect of reducing
$R_o$ significantly but increasing the sideward HBT radius parameter,
$R_s$. Strong transverse flow reduces both transverse HBT radius
parameters at large transverse momentum - in particular $R_o$.
Finally, we used the recent central NA44 $Pb+Pb$ data to extract HBT
radius parameters and to investigate the model dependences of the
transverse sources sizes. The sideward and longitudinal HBT radius
parameters are less sensitive to the details of the freeze-out which
allows us to estimate the initial transverse source size $R_0\sim9-11$
fm and freeze-out time $\tau_f\sim 8-11$ fm/c.  The outward HBT radius
parameter is, however, very sensitive to a number of model dependent
quantities leading to the above mentioned fluctuations.

\section*{Acknowledgements}
We would like to thank Larry McLerran and Scott Pratt
for stimulating discussions.

\end{document}